\documentclass[proceedings, preprint]{rmaa}

\usepackage{paralist}
\usepackage{amsmath}
\usepackage{gensymb}
\usepackage{multirow} 
\usepackage{psfrag,color}

\providecommand{\abs}[1]{\lvert#1\rvert}

\SetYear{2022}
\SetConfTitle{Prospects for low-frequency radio astronomy in S. A.}

\title{Updates on the glitching pulsar monitoring campaign performed from IAR} 

\author{
  E. Zubieta,\altaffilmark{1,2} 
  S. del Palacio,\altaffilmark{2,3}
  F. Garcia,\altaffilmark{2}
  S. B. Araujo Furlan,\altaffilmark{4,5}
  G. Gancio,\altaffilmark{2}
  C. O. Lousto,\altaffilmark{6},
  J. A. Combi,\altaffilmark{7} \\
  \& PuMA Collaboration.}

\altaffiltext{1}{Facultad de Ciencias Astronómicas y Geofísicas, Universidad Nacional de La Plata. Paseo del Bosque s/n, FWA, B1900, Argentina.}

\altaffiltext{2}{Instituto Argentino de Radioastronomía, CONICET--CICPBA--UNLP. Cno. Gral. Belgrano km 40, Pereyra, Argentina.}

\altaffiltext{3}{Department of Space, Earth and Environment, Chalmers University of Technology. SE-412 96 Gothenburg, Sweden.}

\altaffiltext{4}{Instituto de Astronom\'\i{}a Te\'\o{}rica y Experimental, CONICET-UNC, Laprida 854, X5000BGR, Córdoba, Argentina.}

\altaffiltext{5}{Facultad de Matemática, Astronomía, Física y Computación, UNC. Av. Medina Allende s/n , Ciudad Universitaria, CP:X5000HUA, Córdoba, Argentina.}

\altaffiltext{6}{Center for Computational Relativity and Gravitation (CCRG), School of Mathematical Sciences, Rochester Institute of Technology, 85 Lomb Memorial Drive, Rochester, New York 14623, USA.}

\altaffiltext{7}{Departmento de Física, Universidad de Jaén. Campus Las Lagunillas s/n, 23071 Jaén, España.}

\shortauthor{Zubieta et. al.}
\shorttitle{Pulsar monitoring campaign from IAR}

\listofauthors{E. Zubieta, S. del Palacio, F. García, S. B. Araujo Furlan, G. Gancio, C. O. Lousto, J. A. Combi \& PuMA Collaboration}
\indexauthor{E. Zubieta}
\indexauthor{S. del Palacio}
\indexauthor{F. Garcia}
\indexauthor{S. B. Araujo Furlan}
\indexauthor{G. Gancio}
\indexauthor{C. O. Lousto}
\indexauthor{J. A. Combi}

\abstract{Pulsars are known for their exceptionally stable rotation. However, this stability can be disrupted by glitches, sudden increases in rotation frequency whose cause is poorly understood. In this study, we present some preliminary results from the pulsar monitoring campaign conducted at the IAR since 2019. We present measurements from timing solution fits of the parameters of five glitches: one glitch in the Vela pulsar, one in PSR J0742$-$2822, one in PSR J1740$-$3015, and two mini-glitches in PSR J1048$-$5832. Finally, we applied the vortex creep model to characterize the inter-glitch period of Vela. However, the preliminary results yielded highly degenerate and loosely constrained parameters.}

\resumen{Los púlsares cuentan con una rotación excepcionalmente estable. Sin embargo, esta estabilidad puede verse afectada por \textit{glitches}. Este fenómeno consiste en un aumento repentino en la frecuencia de rotación de los púlsares y su causa aún no se encuentra bien comprendida. En este trabajo mostramos algunos resultados preliminares de la campaña de monitoreo de púlsares con observaciones del Instituto Argentino de Radioastronomía (IAR) que comenzó en 2019. Caracterizamos a partir del modelo de \textit{timing} cinco \textit{glitches}: uno en el pulsar de Vela, uno en PSR J0742$-$2822, uno en PSR J1740$-$3015 y dos \textit{mini-glitches} en PSR J1048$-$5832. Por último, aplicamos el modelo de arrastre de vórtices para caracterizar el periodo entre glitches de Vela, pero los resultados preliminares mostraron párametros altamente degenerados y poco restringidos.}
\addkeyword{Pulsars: general}
\addkeyword{Radio continuum: general}
\addkeyword{Method: data analysis}

\begin{document}
% Typeset article header
\maketitle

\section{Introduction}
\label{sec:intro}
%Pulsars are neutron stars with very high moment of inertia, which provides them with exceptionally stable rotation, making them as accurate as an atomic clock \citep{2012MNRAS.427.2780H}. 
Pulsars are neutron stars that present pulsed emission (typically at radio frequencies). Their high moment of inertia provides them with an exceptionally stable rotation, which in some cases makes them as accurate as an atomic clock \citep{2012MNRAS.427.2780H}. 
Typically, the rotation frequency of pulsars decreases regularly due to the loss of angular momentum through electromagnetic emission.
However, their rotation stability can be disrupted by \textit{glitches}. This phenomenon consists of a sudden spin up in the rotation frequency of the pulsar, whose cause is not well understood yet. Glitches were discovered 50 years ago, and now nearly 200 pulsars (mostly young ones) have been observed to exhibit at least one glitch \citep{2011MNRAS.414.1679E, 2017A&A...608A.131F, 2018IAUS..337..197M, 2013MNRAS.429..688Y}. 
Glitches manifest as an abrupt increase in the rotation frequency and are often accompanied by an increment in the spin-down rate. Post-glitch recovery often consists of an exponential decay of the frequency followed by long-term relaxation of the increase in the spin-down rate. In the vortex creep model \citep{2022MNRAS.511..425G}, the post-glitch spin-down rate behavior is believed to mimic the response of the vortex creep regions to a glitch.

Since 2019, the Pulsar Monitoring in Argentina\footnote{https://puma.iar.unlp.edu.ar/} (PuMA) collaboration has been performing high-cadence observations of a set of pulsars \citep{Gancio2020} from the southern hemisphere that have shown glitches before \citep{2022MNRAS.509.5790L}. Observations are performed with the antennas from the Argentine Institute of Radio astronomy (IAR). 
%This monitoring is of great importance, as 
The biggest asset of this monitoring is its high-cadence, which increases the chance of detecting mini glitches and characterizing the post-glitch recovering phase. In particular, we plan to keep monitoring the Vela pulsar hoping to capture a future glitch ``live'' during our 3.5-h daily observations. So far we have detected two mini-glitches on PSR J1048$-$5832 \citep{2023MNRAS.tmp..686Z}, one on PSR J1740$-$3015 \citep{2022ATel15838....1Z}, confirmed one on PSR J0742$-$2822 \citep{2022ATel15638....1Z}, and detected two on the Vela pulsar \citep{2019ATel12482....1L, 2021ATel14806....1S}. Here we present the preliminary results of some of the pulsar glitches we detected so far, and also we apply the vortex creep model on the last Vela glitch to try and estimate the epoch of the next glitch.

\section{Observations and methods}
\label{sec:obs}

\subsection{Observations at the Argentine Institute of Radio astornomy}
\label{sec:IAR}

The Argentine Institute of Radio astronomy is located near the city of La Plata, Argentina. It is equipped with two 30-meter single-dish antennas, ``Carlos M. Varsavsky" and ``Esteban Bajaja", which have the capability to observe a declination range of $-90 \degree < \delta <-10 \degree$ and an hour angle range of two hours east/west: $-2 \mathrm{h} < t < 2 \mathrm{h}$ \citep{Gancio2020}.

We are carrying intensive monitoring of a set of bright glitching pulsars in the southern hemisphere at 1400~MHz. Our observational program involves high-cadence observations, conducted up to daily, with each observation lasting up to 3.5 hours per day. This allows us to build a unique and robust database, capable of detecting and characterizing thoroughly both large and small glitches. 

\subsection{Pulsar timing technique}
\label{sec:timing}

On the one hand, the rotation of pulsars is monitored by recording the times of arrival (ToAs) of the pulses. On the other hand, a mathematical model that characterizes the pulsar's rotation and the propagation of pulses through the interstellar medium, called the timing model, is developed. This model is used to predict the ToAs, and the difference between the observed and predicted ToAs is called the timing residuals. This technique can be used to study diverse physical phenomena, %such as the propagation of gravitational waves, properties of the interstellar medium, and 
including the internal structure of pulsars that gives rise to glitches \citep{2004hpa..book.....L}.

The temporal evolution of a pulsar rotation can be modeled as a Taylor expansion \citep{2011MNRAS.414.1679E}:
\begin{equation}
    \phi(t)=\phi_{0}+\nu_0(t-t_0)+\frac{1}{2}\dot{\nu}_0(t-t_0)^2+\frac{1}{6}\Ddot{\nu}_0(t-t_0)^3 ,
\end{equation}
where $\nu_0$, $\dot{\nu}_0$ and $\Ddot{\nu}_0$ are the frequency and its derivatives. If the model is accurate, residuals should be randomly distributed around zero. However, when a glitch occurs, the pulsar increments its rotation frequency and the pulses begin to arrive before the predicted time, % predicted by the timing model. 
leading to a trend of increasingly more negative residuals (e.g. top panel in Fig.~\ref{fig:Vela}).
Glitches are characterized phenomenologically by an induced jump $\phi_\mathrm{g}$ in the pulsar phase \citep{1987AuJPh..40..725M}:
\begin{equation}\label{ec: glitch-model}
\begin{split}
    \phi_\mathrm{g}=\Delta \phi + \Delta \nu_\mathrm{p} (t-t_\mathrm{g}) + \frac{1}{2} \Delta \dot{\nu}_\mathrm{p} (t-t_\mathrm{g})^2 + \\  \frac{1}{6} \Delta \Ddot{\nu_\mathrm{p}}(t-t_\mathrm{g})^3+ 
    \sum_i [1-\exp{\left(\frac{t-t_\mathrm{g}}{\tau^i_\mathrm{d}}\right)} ]\Delta \nu^i_\mathrm{d} \, \tau^i_\mathrm{d},
\end{split}
\end{equation}
where $\Delta \nu_\mathrm{p}$, $\Delta \dot{\nu}_\mathrm{p}$ and $\Delta \Ddot{\nu_\mathrm{p}}$ are the permanent jump in the frequency and its derivatives. $\Delta \phi$ is an additional phase jump that counteracts the uncertainty on the glitch epoch $t_\mathrm{g}$, and $\Delta \nu_\mathrm{d}^{i}$ are components of the frequency jump that decay after a time $\tau^{i}_\mathrm{d}$, and may or may not be present in the glitch model. In addition, %from these parameters, 
the degree of recovery $Q$ of a glitch is defined as $Q=\Delta \nu_\mathrm{d} / \Delta \nu_\mathrm{g}$, which compares the transient jumps and the total jump in frequency.

The initial parameter sets for the timing models were obtained from the ATNF pulsar catalog \citep{2005AJ....129.1993M} and subsequently updated by ourselves. The timing residuals were calculated with the \texttt{Tempo2} \citep{2006MNRAS.369..655H} software package and we used the \texttt{glitch} plug-in in \texttt{Tempo2} to subdivide the residuals in regions and fit $\nu_0$ and $\dot\nu_0$ in each of them.

\subsection{Vortex creep model}
\label{sec:vortex}

The currently most accepted model to explain glitches in neutron stars is the one proposed by \citet{1969Natur.224..673B}. According to this model, the interior of the neutron star is in a superfluid state, where rotation is achieved through the formation of quantized vortex lines. In the crust of the neutron star, these vortex lines coexist with lattice nuclei. In the vortex creep model \citep{1989ApJ...346..823A}, it is suggested that crustquakes in the crust of the neutron star can lead to the formation of vortex traps and trigger an avalanche of vortex unpinning. The unpinned vortices transfer their excess angular momentum to the crust, resulting in a glitch. Moreover, it is proposed that the crustal breaking may change the magnetospheric characteristics and that parts of the superfluid that have experienced unpinning become decoupled from the external braking torque, affecting the star's braking rate. It is also suggested that the post-glitch relaxation in the rotation of the crust and the superfluid can reflect the response of the superfluid to the changes induced by the glitch. In certain regions of the superfluid, the post-glitch behavior may imitate the response of vortex creep regions, which can be either linear or nonlinear depending on the changes induced by the glitch.

According to this model, in certain parts of the superfluid, vortex creep exhibits a linear relationship with glitch-induced changes, leading to an exponential relaxation:
\begin{equation}\label{ec: exp}
   \Delta \dot{\nu} = - \frac{I_{\mathrm{exp}}}{I_\mathrm{c}} \frac{\Delta \nu_\mathrm{g}}{\tau_\mathrm{exp}} \, e^{-t/\tau_\mathrm{exp}},
\end{equation}
whereas in the non-linear regime:
\begin{equation}\label{ec: lin}
   \Delta \dot{\nu} = \frac{I_{A}}{I_\mathrm{c}} \dot\nu_0 \left( 1 - \frac{1-\dfrac{\tau_\mathrm{nl}}{t_0} \ln{\left[ 1+ \left(e^\frac{t_0}{\tau_\mathrm{nl}}-1\right) e^\frac{-t}{\tau_\mathrm{nl}} \right]}}{1-e^{-t/\tau_\mathrm{nl}}} \right).
\end{equation}
Here, $I_\mathrm{exp}$ and $I_A$ are the total moment of inertia of linear and non-linear creep regions, respectively, and $I_\mathrm{c}$ is the effective moment of inertia of the crust; $\tau_\mathrm{exp}$ and $\tau_\mathrm{nl}$ are the characteristic times of each region, and $t_0$ is the waiting time until glitch-induced changes recover completely (i.e. the theoretical prediction for the inter glitch time).

\section{Detection and analysis of glitches}

\begin{table}[!t]\centering
  \setlength{\tabnotewidth}{\columnwidth}
  \tablecols{4}
  \setlength{\tabcolsep}{0.4\tabcolsep}
  \caption{Glitch parameters obtained from the timing model.}
   \begin{tabular}{llrr}     
     \toprule
     \multirow{ 2}{*}{} PSR & $\mathrm{\LARGE t_g}$ & $\Delta\nu_\mathrm{g} / \nu$  & $\Delta\dot\nu_\mathrm{g} / \dot \nu$ \\ 
     & (MJD) & $(10^{-9})$ &  $(10^{-3})$\\
     \midrule
     J0742$-$2822 & 59839.4(5) & 4294.97(2) & 51.0(7) \\
     J0835$-$4510 & 59417.6193(2) & 1246.9(5) & 84(5)\\
     J1048$-$5832 & 59203.9(5) & 8.89(9) & -0.6(1) \\
     J1048$-$5832 & 59540(2) & 9.9(3) & $\sim$0 \\
     J1740$-$3015 & 59935.1(4) & 323(3) & 1.9(1)\\
     \bottomrule
   \end{tabular}
  \label{tab:glitch-parameters}
 \end{table}

\subsection{PSR J0835$-$4510}
\label{sec:J0835}

\begin{figure}[!t]\centering
  \vspace{0pt}
  \includegraphics[width=0.99\columnwidth]{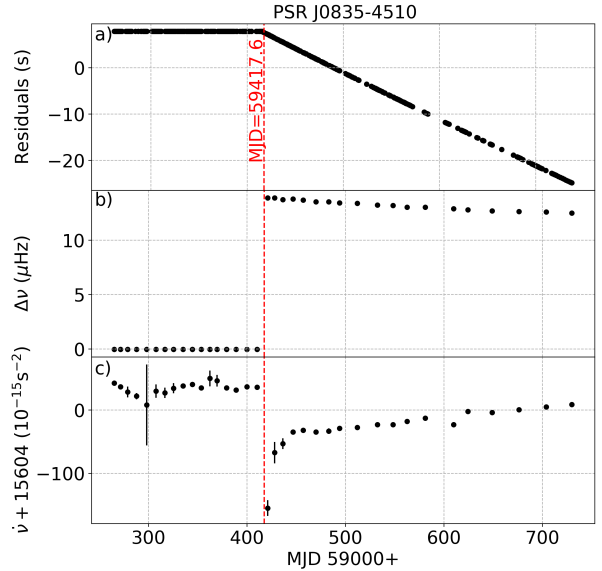}
  \caption{2021 glitch in Vela. \emph{Panel a)}: Residual behavior indicative of a glitch at MJD 59417.6. \emph{Panel b)}: An expanded plot of $\Delta\nu$, where the mean post-glitch value has been subtracted from the post-glitch data. \emph{Panel c)}: Variations in the frequency first derivative, $\Delta\dot{\nu}$. The vertical dashed line denotes the epoch of the glitch.}
  \label{fig:Vela}
\end{figure}

On July 23, 2021, we reported the detection of a new glitch (\#22) in the Vela pulsar. In Fig.~\ref{fig:Vela}a), we show the residuals of Vela before including the glitch in the timing model. Fig.~\ref{fig:Vela}b) and Fig.~\ref{fig:Vela}c) present the evolution of $\Delta\nu$ and $\Delta\dot{\nu}$. We measured a magnitude of the glitch of $\Delta\nu / \nu \approx 1.25\times 10^{-6}$. The parameters of the glitch corresponding to the permanent jump are listed in Table \ref{tab:glitch-parameters}. 
In Fig.~\ref{fig:Vela}c) we also observe an increment of $\abs{\dot{\nu}}$ together with an exponential decay. Our high-cadence monitoring allowed us to characterize thoroughly the transitory components of the glitch. We found two exponential decays, $\tau_\mathrm{d1} = 6.400(2)~\mathrm{d}$ and $\tau_\mathrm{d2} = 0.994(8)~\mathrm{d}$, together with a magnitude of $\Delta\nu_\mathrm{d1} = 3.15(12)\times 10^{-8}~\mathrm{s}^{-1}$ and $\Delta\nu_\mathrm{d2} = 9.9(6)\times 10^{-8}~\mathrm{s}^{-1}$. These transient jumps correspond to $Q_1=0.2(1)\%$ and $Q_2=0.7(1)\%$, which indicates that the glitch is dominated by the permanent jumps.

\begin{figure}[t]\centering
  \vspace{0pt}
  \includegraphics[width=0.99\columnwidth]{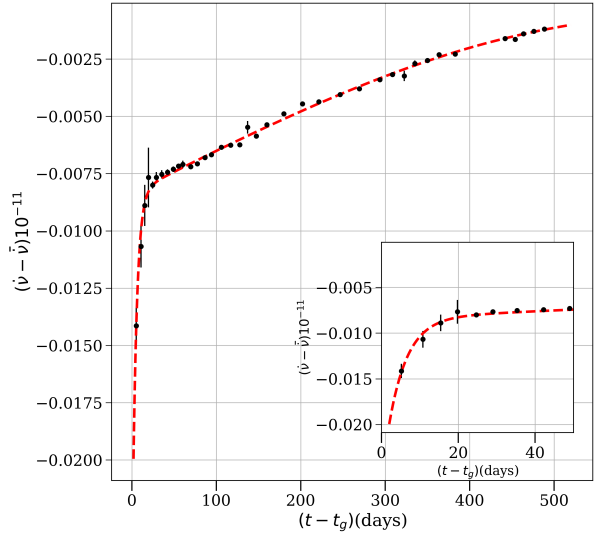}
  \caption{Post-glitch behavior of $\dot\nu$ following the vortex creep model. The inset highlights the fit to the data close to the time of the glitch, where the exponential-decay terms predominate.}
  \label{fig:fit}
\end{figure}

The post-glitch relaxation can be interpreted in terms of the vortex creep model given that $\dot \nu$ shows an exponential relaxation followed by a linear recovery. Thus, we fitted $\dot \nu$ using Eqs.~(\ref{ec: exp})--(\ref{ec: lin}) with the inclusion of two decay terms in the model, in consistency with the timing analysis. However, the fitting yielded loosely constrained and highly degenerated parameters, likely due to insufficient precision in the data. Specifically, we obtained a value of $t_0 = 421(89)$~d, which is inconsistent with the inter-glitch time of the Vela pulsar ($\sim$2--3~yr). Nevertheless, a more detailed analysis of the dataset (including further cleaning of the observations) will be addressed to corroborate these preliminary results and also search for putative mini-glitches around that epoch. 

\subsection{PSR J1048$-$5832}
\label{sec:J1048}

\begin{figure}[!t]\centering
  \vspace{0pt}
  \includegraphics[width=0.99\columnwidth]{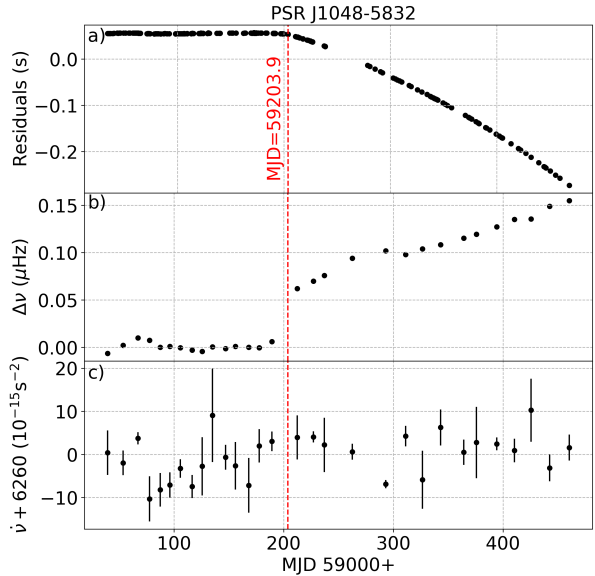}
  \caption{The same as Fig.~\ref{fig:Vela} but for the 2020 glitch in PSR J1048$-$5832.}
  \label{fig:J1048_G1}
\end{figure}

\begin{figure}[!t]\centering
  \vspace{0pt}
  \includegraphics[width=0.99\columnwidth]{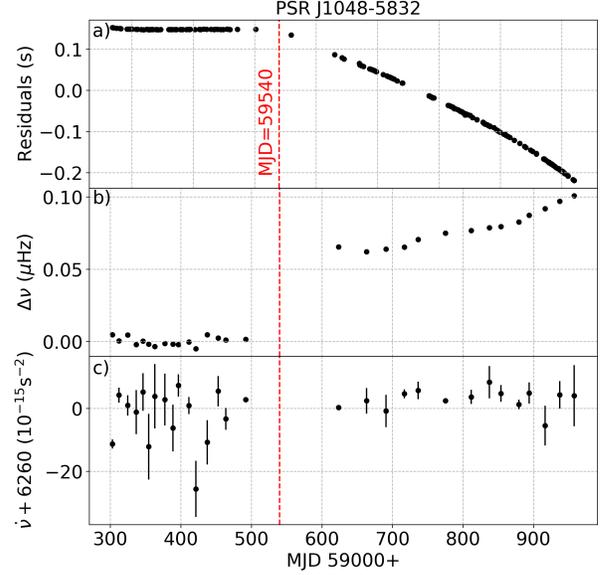}
  \caption{The same as Fig.~\ref{fig:Vela} but for the 2021 glitch in PSR J1048$-$5832.}
  \label{fig:J1048_G2}
\end{figure}

Between 1992 and 2014, seven glitches were reported in this pulsar. We identified two new small ones on MJD 59203.9(5) (December 20, 2020) and MJD 59540(2) (November 22, 2021). The residuals indicative of the glitches are shown in Fig.~\ref{fig:J1048_G1}a) and Fig.~\ref{fig:J1048_G2}a). Fig.~\ref{fig:J1048_G1}b) and Fig.~\ref{fig:J1048_G2}b) show the behavior of the frequency around each glitch, and Fig.~\ref{fig:J1048_G1}c) and Fig.~\ref{fig:J1048_G2}c) present the evolution of $\Delta\dot{\nu}$. The permanent jumps corresponding to both glitches are shown in Table \ref{tab:glitch-parameters}. 
These are the smallest glitches detected so far for this pulsar, and no exponential decay was observed in any of them.

\subsection{PSR J0742$-$2822}
\label{sec:J0742}

\begin{figure}[!t]\centering
  \vspace{0pt}
  \includegraphics[width=0.99\columnwidth]{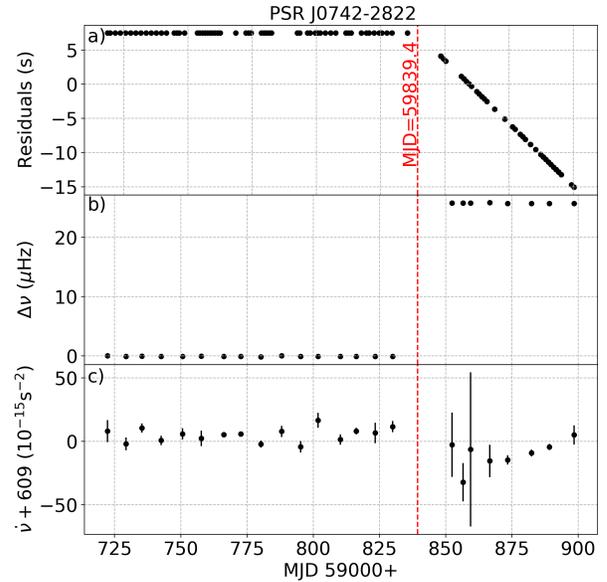}
  \caption{The same as Fig.~\ref{fig:Vela} but for the 2022 glitch in PSR J0742$-$2822.}
  \label{fig:J0742}
\end{figure}

\citet{2022ATel15622....1S} reported the ninth glitch in this pulsar on MJD 59839.4 (September 21, 2021). We corroborated the glitch through the residuals shown in Fig.~\ref{fig:J0742}a), which has a magnitude of $\Delta\nu/\nu \approx 4.3\times 10^{-6}$. Our values found for the permanent jumps of the frequency and its derivative are reported in Table \ref{tab:glitch-parameters},
and their behavior are shown in Figs.~\ref{fig:J0742}b), c). No signs of exponential recovery were found, probably because of the lack of observations within the next $\sim15$ days after the glitch epoch. Figure~\ref{fig:J0742}c) shows that the post-glitch spin-down rate $\abs{\dot\nu}$ is larger than the pre-glitch one.

\subsection{PSR J1740$-$3015}
\label{sec:J1740}

\begin{figure}[!t]\centering
  \vspace{0pt}
  \includegraphics[width=0.9\columnwidth]{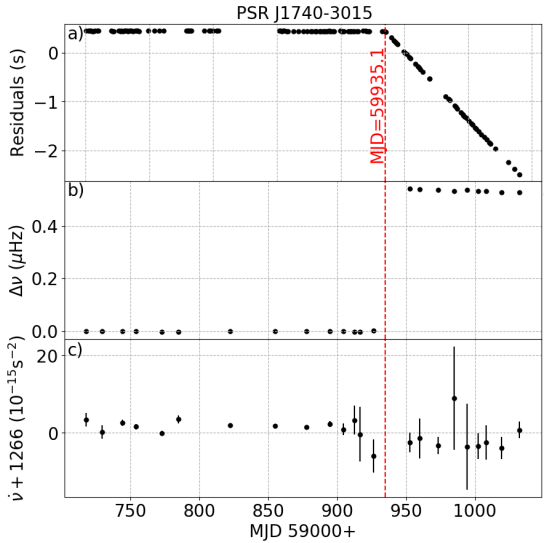}
  \caption{The same as Fig.~\ref{fig:Vela} but for the 2022 glitch in PSR J1740$-$3015.}
  \label{fig:J1740}
\end{figure}

PSR J1740$-$3015 is a frequently glitching pulsar, with already 37 glitches reported\footnote{https://www.atnf.csiro.au/people/pulsar/psrcat/glitchTbl.html}. We detected another glitch in this pulsar on  MJD 59935.1(4) (December 22, 2022), with a magnitude of $\Delta\nu/\nu \approx 3.23\times 10^{-7}$. The residuals indicating the glitch are shown on Fig.~\ref{fig:J1740}a), and the values for the permanent jumps of the frequency and its derivative are reported in Table~\ref{tab:glitch-parameters}.
Figures~\ref{fig:J1740}b), c) present the evolution of $\Delta\nu$ and $\Delta\dot{\nu}$. There was an increment in the spin-down rate $\abs{\dot\nu}$ after the glitch but no signal of exponential recovery was found.

\section{Conclusions}
\label{sec:conclusions}

We are currently carrying an intensive monitoring program of bright pulsars in the southern hemisphere using both IAR radiotelescopes. This program has yielded encouraging results in the detection of glitches, demonstrating that significant contributions to pulsar timing studies can be made with these instruments. In particular, our high-cadence observations allowed us to detect two small glitches in PSR J1048$-$5832 and to perform a thorough analysis of the 2021 glitch in the Vela pulsar, which revealed two decay terms. Nevertheless, the application of the vortex creep model to such a glitch resulted in loosely constrained parameters. This analysis will undergo further re-processing of observations in the future.

\end{document}